\documentclass{PoS}
\usepackage{amsmath,amstext,amsfonts,amsbsy,amssymb,amscd,bbm,epsfig,lscape}

\newcommand{\ba}{\begin{array}}
\newcommand{\ea}{\end{array}}

\newcommand{\Id}{\mathbf{1}}

\newcommand{\Dslash}{\relax{\kern+.25em / \kern-.70em D}}

\newcommand{\fm}{\rm fm}
\newcommand{\MeV}{\rm MeV}
\newcommand{\GeV}{\rm GeV}

\newcommand{\Real}{\relax{\mathsf{\Gamma\kern-.35em R}}}
\newcommand{\Int}{\relax{\mathsf{Z\kern-.40em Z}}}

\newcommand{\SUt}{\mbox{SU}(3)}

\newcommand{\half}{{\scriptstyle{{1\over 2}}}}

\newcommand{\ihalf}{{\scriptstyle{{i\over 2}}}}

\newcommand{\MSbar}{{\overline{\rm MS}}}

\newcommand{\gbar}{\kern1pt\overline{\kern-1pt g\kern-0pt}\kern1pt}
\newcommand{\mbar}{\kern2pt\overline{\kern-1pt m\kern-1pt}\kern1pt}
\newcommand{\obar}[1]{\kern3pt\overline{\kern-2pt #1\kern-0pt}\kern1pt}

\newcommand{\mren}[1]{m_{{\rm R} #1}}
\newcommand{\muren}[1]{\mu_{{\rm R} #1}}

\newcommand{\hopc}{\kappa_{\rm c}}

\newcommand{\fX}{f_{\rm\scriptscriptstyle X}}
\newcommand{\fP}{f_{\rm\scriptscriptstyle P}}

\newcommand{\fA}{f_{\rm\scriptscriptstyle A}}
\newcommand{\fV}{f_{\rm\scriptscriptstyle V}}

\newcommand{\ZA}{Z_{\rm\scriptscriptstyle A}}
\newcommand{\ZV}{Z_{\rm\scriptscriptstyle V}}

\newcommand{\Oa}{\mbox{O}(a)}
\newcommand{\Oasq}{\mbox{O}(a^2)}

\newcommand{\icsw}{c_{\rm sw}}

\newcommand{\icA}{c_{\rm\scriptscriptstyle A}}

\newcommand{\abar}{\kern1pt\overline{\kern-1pt a\kern-0.5pt}\kern1pt}

\title{Twisted mass QCD for weak matrix elements\thanks{CERN-PH-TH/2006-203}}

\ShortTitle{tmQCD for weak matrix elements}

\author{\speaker{Carlos Pena}\\
        CERN, Physics Department, Theory Division, CH-1211 Geneva 23, Switzerland \\
        E-mail: \email{carlos.pena.ruano@cern.ch}}

\abstract{I report on the application of tmQCD techniques to the computation
of hadronic matrix elements of four-fermion operators. Emphasis is put on the 
computation of $B_K$ in quenched QCD performed by the ALPHA Collaboration. The 
extension of tmQCD strategies to the study of neutral $B$-meson mixing is briefly 
discussed. Finally, some remarks are made concerning proposals to apply 
tmQCD to the computation of $K\to\pi\pi$ amplitudes.}

\FullConference{XXIV International Symposium on Lattice Field Theory\\
		 July 23-28 2006\\
		 Tucson Arizona, US}

\begin{document}

\section{Introduction: Flavour Physics vs lattice systematics}

During the last five years, a new generation of experimental facilities dominated
by B factories have brought Flavour Physics to the era of precision studies.
Uncertainties on decay and mixing amplitudes of kaons, $D$- and $B$-mesons
have plummeted, setting more stringent constraints than ever on the accuracy
required for theoretical estimates aimed at determining Standard Model (SM)
parameters or probing new physics. Indeed, although a determination of CKM
matrix elements essentially free from hadronic uncertainties starts to be possible
\cite{Bona:2006ah}, knowing all the relevant matrix elements to the required precision
is still essential in order to check the consistency of SM predictions and
set bounds on effects beyond the SM.

The requirement of few percent accuracy demands, in particular, a fully
first-principles approach to the long-distance regime of strong interactions
in which all the systematics is consciously brought under control. Mandatory
features include:

\begin{itemize}

\item Dynamical simulations with at least $2(\mbox{{\em light}})+1$
flavours of sea quarks.

\item Good conceptual control over the regularisation.

\item Good control of all the symmetries (especially: flavour symmetries).

\item Fully non-perturbative renormalisation of all the composite operators involved.

\item Elimination of cutoff dependences.

\end{itemize}

Wilson and chirally symmetric fermions are preferred on conceptual grounds.
Wilson fermions have disadvantages from the point of view of flavour symmetries
and treatment of ultraviolet cutoff dependencies, but are already able to
approach the light quark regime in dynamical simulations
\cite{LG,DelDebbio:2006cn,Jansen:2006rf,Gockeler:2006ns,Gockeler:2006vi},
although recent progress with
dynamical chiral fermions has proved equally impressive (see \cite{Kaneko:2006pa}
and references therein).
Twisted mass QCD (tmQCD) is a variant of the Wilson regularisation that
potentially allows for a better control of chiral symmetry breaking and
cutoff effects, which turns particularly advantageous in the computation
of weak matrix elements. In this context, it may offer a convenient compromise
between an adequate control over chiral symmetry and numerical affordability.
Precision quenched computations, that properly deal with all the systematics
apart from dynamical quark effects, are an essential step in order to understand
these issues and prepare the terrain for dynamical studies.
Not least importantly, they offer sound arguments for the choice of
regularisation both for sea quarks and for the valence sector of
mixed action approaches.

This paper deals mainly with quenched numerical results for weak matrix elements 
obtained from tmQCD. In Section 2 the ALPHA Collaboration computation
of $B_K$ in tmQCD \cite{Dimopoulos:2006dm} is discussed. Section 3
briefly reviews the extension of the strategy to deal with neutral $B$-meson
mixing amplitudes, and reports on the project status. Section 4 deals with tmQCD 
proposals to study $K\to\pi\pi$ amplitudes. Finally, in Section 5 some final remarks 
are made.

Much of the work presented \cite{Dimopoulos:2006dm,Guagnelli:2005zc,Palombi:2005zd,Palombi:2006pu} is part of the ALPHA Collaboration
research programme. General reviews of progress in kaon and $B$-physics on
the lattice have been provided at this conference by W.~Lee and T.~Onogi
\cite{Lee:2006cm,Onogi}.

\section{$B_K$ in quenched (tm)QCD}

\subsection{Lattice QCD and indirect CP violation in kaon decays}

Indirect CP violation in $K\to\pi\pi$ decays is measured by the parameter
$\varepsilon_K$, defined in terms of kaon decay amplitudes as
\begin{gather}
\varepsilon_K = \frac{T(K_L\to(\pi\pi)_{I=0})}{T(K_S\to(\pi\pi)_{I=0})} \, ,
\end{gather}
where $I$ is the total isospin of the two-pion state. Experiment yields
\cite{Yao:2006px}
\begin{gather}
\varepsilon_K = [2.232(7) \times 10^{-3}]\,\,e^{i\phi_\varepsilon} \,,~~~~
\phi_\varepsilon = (43.5 \pm 0.7)^\circ \,.
\end{gather}

At leading order in an Operator Product Expansion (OPE) treatment of electroweak 
interactions, the Standard Model (SM) prediction for $|\varepsilon_K|$ can be
written as \cite{Battaglia:2003in}
\begin{gather}
|\varepsilon_K| \simeq C_\varepsilon \hat B_K
{\rm Im}\{V_{td}^*V_{ts}\}
\left\{
{\rm Re}\{V_{cd}^*V_{cs}\}\left[\eta_1 S_0(x_c)-\eta_3 S_0(x_c,x_t)\right] -
{\rm Re}\{V_{td}^*V_{ts}\}\eta_2 S_0(x_t)
\right\} \, .
\end{gather}
Here $C_\varepsilon = G_{\rm F}^2 F_K^2 M_K M_W^2/(6\sqrt{2}\pi^2\Delta M_K)$,
$S_0(x_t)$ and $S_0(x_c,x_t)$ ($x_i=m_i^2/M_W^2$) parameterise the Wilson
coefficients of the OPE, $\eta_{1,2,3}$ are short-distance QCD corrections
to the latter (known to NLO), and
\begin{gather}
\hat B_K = \frac{\langle \bar K^0|\hat O^{\Delta S=2}| K^0 \rangle}{\frac{8}{3}F_K^2 M_K^2} \, ,
\end{gather}
where $\hat O^{\Delta S=2}$ is the effective four-quark interaction operator
\begin{gather}
O^{\Delta S=2} = (\bar s \gamma_\mu^{\rm L} d)(\bar s \gamma_\mu^{\rm L} d) \, ,
\end{gather}
$\gamma_\mu^{\rm L}=\gamma_\mu(\Id-\gamma_5)$,
and the hat denotes renormalisation group invariant (RGI) matrix elements. The
dimensionless parameter $\hat B_K$ thus provides the long-distance, non-perturbative
QCD contribution, and largely dominates the uncertainty on the SM value for 
$|\varepsilon_K|$.
In the standard Unitarity Triangle (UT) analysis of CP violation in the SM, the value of $|\varepsilon_K|$ provides a hyperbola in the $(\bar\rho,\bar\eta)$ plane. After the recent generation of experimental
results from $B$-factories, this is one of the least precise UT constraints.
Improving the accuracy of $\hat B_K$ is hence essential in order
to derive stringent bounds on the amount of non-SM CP violation in kaon decay.

Besides quenching, which is an uncontrolled
source of systematic error, the most important source of
uncertainty in lattice QCD computations of $B_K$ with Wilson fermions 
arises from operator
renormalisation. In standard notation, the operator $O^{\Delta S=2}$ is customarily split into parity-even and parity-odd parts as
\begin{gather}
O^{\Delta S=2} = O_{\rm VV+AA} - O_{\rm VA+AV} \,.
\end{gather}
Since parity is a QCD symmetry, the only
contribution to the $K^0$--$\bar K^0$ matrix element comes from $O_{\rm VV+AA}$.
In regularisations which respect chiral symmetry,
the latter operator is multiplicatively
renormalisable. If chiral symmetry is not preserved,
$O_{\rm VV+AA}$ mixes with four other dimension-6
operators~\cite{Marti:4fPT,Bernardetal:4fPT,Bernard:lat87,
Guptaetal:bk,Doninetal:4fRIMOM} with positive parity:
\begin{eqnarray}
(O_{\rm R})_{\rm VV+AA}(\mu) = Z_{\rm VV+AA}(g_0, a\mu) \Big [ O_{\rm VV+AA} (g_0)
+ \sum_{i = 1}^4 \Delta_i(g_0) O_i(g_0) \Big ]
\end{eqnarray}
The operators $ O_i(g_0)$ belong to different chiral
representations than $O_{\rm VV+AA}$. The mixing coefficients
$\Delta_i(g_0)$ are finite functions of the bare coupling,
while the renormalisation constant $Z_{\rm VV+AA}$
diverges logarithmically in $a\mu$.

Two proposals have attempted to eliminate operator mixing.
They are both based on the 
observation~\cite{Bernard:lat87,Doninetal:4fRIMOM}
that, even in the absence of chiral symmetry, the
operator $O_{\rm VA+AV}$ is protected from finite operator mixing
by discrete symmetries, and thus it renormalises multiplicatively, viz.
\begin{eqnarray}
(O_{\rm R})_{\rm VA+AV}(\mu) = Z_{\rm VA+AV}(g_0, a\mu) O_{\rm VA+AV} (g_0) \,.
\end{eqnarray}
The first proposal~\cite{BK:WI} consists in obtaining the physical
$K^0 - \bar K^0$ matrix element of $O_{\rm VV+AA}$ from
a correlation function of the renormalised operator
$O_{\rm VA+AV}$, related to it through axial Ward identities. The method has
been put to test in ref.~\cite{SPQR:bk}, with the result that the $B_K$
estimate turned out to be compatible with the result of
computations that involve operator subtractions.
Unfortunately, the  correlation function of $O_{\rm VA+AV}$ is a
four-point function, while the matrix element of $O_{\rm VV+AA}$
can be extracted from a three-point function. Thus,
the conclusion of \cite{SPQR:bk} is that
this method is successful in eliminating an important source
of systematic errors (operator subtraction) at the cost of
increased statistical fluctuations.

In the work under consideration here, the second proposal~\cite{tmqcd:pap1},
based on twisted mass QCD, is implemented. In tmQCD the breaking pattern of
flavour symmetries is controlled by the value of the twist angle; in
particular, the latter can be tuned so as to preserve part of the axial
subgroup, at the price of breaking vector symmetries,
as well as parity. It is thus possible to set up regularisations
in which the renormalisation of composite operators is greatly
simplified. The relevant case for us is the renormalised
$\langle \bar K^0 \vert O_{\rm VV+AA} \vert K^0 \rangle$
matrix element, which via the tmQCD formalism can be extracted from
a three-point correlation function of the operator $O_{\rm VA+AV}$.
As the tmQCD action differs from the standard Wilson fermion action by a
soft term, the renormalisation properties of composite operators
in mass independent renormalisation schemes are not modified.
In particular, $O_{\rm VA+AV}$ remains multiplicatively renormalisable,
with the same renormalisation constant and running as with
Wilson fermions. Thus finite subtractions are avoided
in the tmQCD determination of $B_K$.

An obvious alternative to avoid renormalisation problems consists
in using regularisations with exact chiral symmetry. However, the computational
costs involved make it difficult to perform continuum limit extrapolations and
study finite volume effects.\footnote{For a state-of-the-art determination
of $\Delta S=2$ matrix elements, see~\cite{Babich:2006bh}.}
In the case of staggered fermions, apart from the operator mixing
(the details of which depend on the specific setup), some
additional problems are present --- large scaling violations unless high
levels of $\Oasq$ improvement are implemented, uncertainties related to
the choice of interpolating operator, as well as other difficulties
related to the breaking of flavour symmetries and the presence of unphysical
flavours.\footnote{See \cite{Lee:2006cm} for an updated discussion of
staggered quark results.}
Wilson fermions therefore offer, {\em a priori}, a good compromise between good
control of the field-theoretical aspects of the problem and affordable
computational costs.

\subsection{tmQCD setup}

We will employ two different fermion actions, namely
\begin{align}
 \label{tmQCD_action2}
 S_{\rm F}^{(\pi/2)} &= a^4 \sum_x \,\, [ \bar{\psi}(x) (D_{\rm w,sw} 
+ m_l + i\gamma_5 \tau^3 \mu_l )\psi(x) \,+\,
\bar s(x) (D_{\rm w,sw} + m_s ) s(x) ] \,,\\
 \label{tmQCD_action4}
 S_{\rm F}^{(\pi/4)} &= a^4 \sum_x \,\, [ \bar u(x) (D_{\rm w,sw} + m_u ) u(x) \,+\,
 \bar{\psi}(x) (D_{\rm w,sw} 
+ m_l + i\gamma_5 \tau^3 \mu_l )\psi(x)
]\,.
\end{align}
The labels on the two actions refer to the values that will eventually be
set for the twist angle.
In Eq.~(\ref{tmQCD_action2}) $\psi=(u,d)^T$, while in Eq.~(\ref{tmQCD_action4}) $\psi=(s,d)^T$. In both cases, the matrix $\tau^3$ acts on flavour space,
and $D_{\rm w,sw}$ is the Wilson-Dirac operator with a Sheikholeslami-Wohlert
term; $\icsw$ is tuned to its non-perturbative value \cite{Luscher:1996ug}. In Eq.~(\ref{tmQCD_action4}) it has been assumed {\em a priori} that the
$s$ and $d$ quarks have degenerate physical masses; while this is not necessary
as long as this action is used in quenched QCD, all the computations
carried out with it are performed in that limit. The action in Eq.~(\ref{tmQCD_action2}), on the other hand, is perfectly well suited for an
unquenched computation, and it has been used to explore the effect of having
non-degenerate $s$ and $d$ quark masses (see below).

The properties of tmQCD have been extensively discussed in several publications
(see \cite{tmqcd:pap1,Shindler:2005vj,Frezzotti:2004pc} and references therein). Here we just remind some basic facts. The physical renormalised masses of twisted quarks and the twist angle $\alpha$ are given by
\begin{align}
M_{\rm R} &= \sqrt{m_{\rm R}^2 + \mu_{\rm R}^2} \,,\\
\tan\alpha &= \frac{\mu_{\rm R}}{m_{\rm R}} \,,
\end{align}
where $m_{\rm R}$ (resp. $\mu_{\rm R}$) are the renormalised standard (twisted)
quark masses.
In order to tune the twist angle to some prescribed value up to $\Oasq$ corrections,
we employ the formulae for the construction of $\Oa$ improved renormalised
masses
\begin{align}
\label{eq:mren}
\mren{,l} &= Z_m [ m_{q,l} ( 1 + b_m a m_{q,l} ) + \tilde b_m a \mu_l^2] \,,\\
\label{eq:muren}
\muren{,l} &= Z_\mu \mu_l ( 1 + b_\mu a m_{q,l} ) \,,
\end{align}
where $m_{q,l}=\half(1/\kappa-1/\hopc)$ is the subtracted bare standard quark mass.

In the case of the $\pi/2$ regularisation, in order to have $\alpha=\pi/2$ it
is enough to set $\mren{,l}$ to zero, which is achieved by setting
\begin{gather}
a  m_{q,l} = - \tilde b_m ( a  \mu_l )^2 \,.
\end{gather}
The $\pi/4$ case is somewhat less trivial. Setting $\alpha=\pi/4$
requires $\mu_{{\rm R},l} = m_{{\rm R},l}$, which via
Eqs. (\ref{eq:mren},\ref{eq:muren}) translates into
\begin{gather}
a m_{q,l} = \dfrac{1}{Z Z_A} a \mu_l \big \{ 1 +
\big [ \dfrac{1}{Z Z_A} (b_\mu - b_m ) - Z Z_A \tilde b_m \big ]
a \mu_l \big \}
\end{gather} 
with $Z \equiv Z_m / (Z_\mu Z_A)$. For a given choice
of $a \mu_l$, $\kappa$ is tuned so that 
$a m_{q,l}$ satisfies one of the two above relations, taking the values of
$\hopc$ and all
the renormalisation constants and improvement
coefficients involved as input. The precision to which the latter are known
poses an implicit constraint on the accuracy of the tuning of the twist angle.

Contact with QCD is made via the change of fermion variables
\begin{gather}
 \label{eq:chi_rot}
 \psi \to \psi' = R(\alpha) \psi, \qquad
 \bar\psi \to \bar\psi' = \bar\psi R(\alpha)\ ,
\end{gather}
where $R(\alpha) = \exp\left\{\ihalf\gamma_5 \alpha\tau^3 \right\}$
and $\psi$ is the twisted quark doublet. This axial rotation induces
a mapping between composite operators in tmQCD and QCD, which is realised
at the level of renormalised correlation functions (or, alternatively,
renormalised matrix elements). The relation we are most interested in is
\begin{gather}
\langle K^0 \vert \,\, (O_R)_{\rm VV+AA} \,\, \vert \bar K^0 \rangle_{\rm QCD} =
- i \langle  K^0 \vert \,\, (O_R)_{\rm VA+AV} \,\, \vert \bar K^0 \rangle_{\rm tmQCD} \,,
\end{gather}
which holds in the continuum limit for the two versions of tmQCD under
consideration. From this identity, $B_K$ can be extracted from a $K^0$--$\bar K^0$
matrix element of the multiplicatively renormalisable operator $O_{\rm VA+AV}$.

It is important to stress that none of the above setups leads to a computation
of $B_K$ that involves fully twisted quarks only. Hence, the automatic $\Oa$
improvement argument of Frezzotti and Rossi \cite{Frezzotti:2003ni} does not apply, and
in order to have full $\Oa$ improvement of the matrix element it would be
necessary to subtract a number of dimension-seven counterterms from the
four-fermion operator. Such a procedure is highly impractical, and has
not been pursued. Hence, leading cutoff effects in $B_K$
are expected to be linear in $a$.

\subsection{Renormalisation}

The non-perturbative renormalisation of the operator $O_{\rm VA+AV}$ has
been addressed in \cite{Guagnelli:2005zc,Palombi:2005zd} using standard Schr\"odinger Functional
(SF) techniques (see e.g. \cite{Sommer:2002en}). After having defined suitable SF intermediate renormalisation
schemes, a recursive step-scaling procedure allows to compute to high accuracy
the renormalisation group (RG) running of the operator in quenched QCD in the continuum limit from a low-energy reference scale $(2L_{\rm max})^{-1}$
close to $\Lambda_{\rm QCD}$ to scales of $O(100~\GeV)$, where reliable contact
with perturbation theory can be made. Together with the renormalisation
constants at $(2L_{\rm max})^{-1}$, this provides RGI renormalisation
factors free from any uncontrolled systematic uncertainty.

\begin{figure}
\vspace{-20truemm}
\center\includegraphics[width=0.6\textwidth]{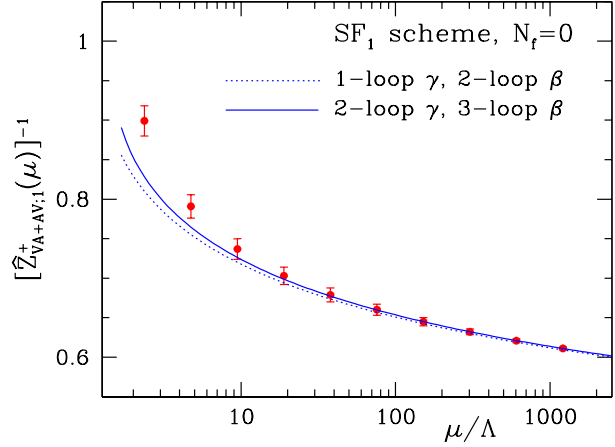}
\vspace{-15truemm}
\caption{RG running $O_{\rm VA+AV}$ in quenched QCD in the SF scheme.
Non-perturbative values 
(circles) are compared to perturbation theory predictions.}
\label{fig1}
\end{figure}

In the particular case of $O_{\rm VA+AV}$, nine different SF schemes were
defined and found to provide consistent results for RGI renormalisation factors. The continuum limit of the RG running was controlled by
performing independent simulations with two different fermion actions
(plain and $\Oa$ improved Wilson fermions). The quality of the result
is illustrated by Figure~\ref{fig1}. This approach is currently being pursued
in order to extend the non-perturbative renormalisation of
$O_{\rm VA+AV}$ to $N_f=2$ QCD \cite{Dimopoulos:2006es}. It is also worth
mentioning that the scope of \cite{Guagnelli:2005zc,Palombi:2005zd} goes well beyond the case
of the $\Delta S=2$ effective Hamiltonian. For instance, \cite{Dimopoulos:2006ma} 
made use of the results in \cite{Guagnelli:2005zc} to address the renormalisation
of  the $\Delta S=1$  effective Hamiltonian with an active charm quark constructed 
with overlap fermions. To that purpose, the logarithmically divergent
renormalisation constants required have been computed through
a matching of non-perturbatively renormalised RGI tmQCD matrix elements to bare 
overlap matrix elements at a reference mass $m_{\rm PS} \simeq m_K$. This procedure
is similar to the one employed in \cite{Hernandez:2001yn}
for the renormalisation of the quark condensate.

\subsection{Simulations for bare matrix elements and systematics}

The bare value of $B_K$ can be extracted from the ratio of SF correlation
functions
\begin{gather}
\label{eq:bare_ratio}
R(x_0) = \frac{3}{16}\frac{-if_{\rm VA+AV}(x_0)}
{[\ZA(\fA(x_0)+\icA a\partial_0\fP(x_0))-i\ZV\fV(x_0)]
[\ZA(\fA'(x_0)+\icA a\partial_0\fP'(x_0))-i\ZV\fV'(x_0)]} \,,
\end{gather}
where $f_{\rm VA+AV}$ is the correlation function of $O_{\rm VA+AV}$ with
two pseudoscalar SF boundary sources and $\fX$ ($X=A_0,P,V_0$) is a
two-point function of the bilinear operator $X$ with a pseudoscalar SF boundary
source. Precise definitions can be found in \cite{Dimopoulos:2006dm}. The combination
of currents in the denominator corresponds to the physical, $\Oa$ improved,
renormalised axial current via the chiral rotation in Eq.~(\ref{eq:chi_rot}).

Quenched simulations have been performed at $\beta=6.0,6.1,6.2,6.3$ for the $\pi/2$
regularisation and $\beta=6.0,6.1,6.2,6.3,6.45$ for $\pi/4$. The physical
masses of the $s$ and $d$ quarks have always been kept degenerate, save for
a subset of simulations meant to probe $\SUt$ flavour breaking effects (see below).
In the $\pi/2$ case, since the $s$ quark is untwisted, it is impossible to reach
pseudoscalar masses in the region of $m_K$, due to
the presence of exceptional configurations. Therefore, pseudoscalar
masses larger than $m_K$ are simulated, and the results are hence extrapolated to
the physical kaon mass. In the $\pi/4$ case, on the contrary, values at
$m_K$ can be obtained by interpolation.\footnote{The $\beta=6.45$ data,
however, have been obtained at larger masses and then extrapolated to
$m_K$.}
The necessary renormalisation constants and improvement coefficients, as well as the 
values of $\hopc$, have been gathered from the literature.\footnote{See Appendix~A
of \cite{Dimopoulos:2006dm} and references therein.} The scale is always fixed via 
the ratio $r_0/a$ as given by \cite{Necco:2001xg}, with $r_0=0.5~\fm$.

The results have been subjected to a number of checks, meant to assess various
systematic uncertainties:

\begin{itemize}

\item {\em Finite volume effects.} In the $\pi/2$ case, simulations at $\beta=6.0$
have been performed for physical lattice sizes around $L \approx 1.5~\fm$ and
$L \approx 2~\fm$, at the lowest pseudoscalar meson mass available. 
Consistent values for all the relevant observables are obtained within errors.
Simulations were hence carried out on lattices with $L \sim 1.5~\fm$.
In the $\pi/4$, a similar study was performed at $\beta=6.0,6.2$ and masses
around $m_K$. In this case the conclusion is that lattice sizes $L \sim 2~\fm$ are
needed to avoid finite volume effects.

\item {\em $\SUt$ breaking effects.} Physical $\SUt$ breaking effects on $B_K$
have been studied in the $\pi/2$ case at $\beta=6.0$. To this purpose, simulations
were performed for three values of the ratio $\epsilon=(M_s-M_d)/(M_s+M_d)$
(where $M_f$ is the physical quark mass), namely $\epsilon=0.00,0.16,0.41$,
at fixed $r_0 m_{\rm PS}=1.78$. No effect was observed on $B_K$ within uncertainties, 
hinting at
small $\SUt$ breaking effects. It has to
be stressed, however, that the simulated pseudoscalar mass is relatively high.

\item {\em Spurious $\SUt$ breaking.} The $\Oasq$ breaking of
vector flavour symmetries induced by the presence of the twisted mass term
has received considerable attention in the literature (see \cite{Shindler:2005vj}). In order
to check its effect, one can compare the pseudoscalar meson mass obtained
for various flavour combinations of twisted and untwisted quarks in the
$\pi/2$ case. In the $M_s=M_d$ limit, the resulting states can be interpreted
as belonging to a multiplet of pseudo-Goldstone bosons; hence, deviations from unity
in the ratios $(m_{\rm PS}/m'_{\rm PS})^2$ of squared pseudoscalar masses
in different flavour channels quantify the
$\Oasq$ vector symmetry breaking. The values of these ratios
show that the effect is never beyond the few percent level, and converges to zero
in the continuum limit. The splitting tends to grow mildly as the quark
mass is decreased. These findings contrast, but are by no means
incompatible, with the observation of much larger amounts of
$\Oasq$ flavour breaking at lighter quark masses \cite{Becirevic:2006ii}.

\end{itemize}

After the publication of \cite{Dimopoulos:2006dm}, a more detailed
analysis of the accuracy of the tuning of quark masses and twist angles
was performed. The
quality of the tuning was found to be satisfactory in all cases save for
the simulations at $\beta=6.1$, mainly in the $\pi/4$ case. This is signalled e.g. by
relatively large differences between the value of the target twist angle, set
to $\pi/2$ or $\pi/4$ when tuning the quark masses via Eqs.~(\ref{eq:mren},\ref{eq:muren}), and the
value obtained by computing the ratio $\muren{,l}/\mren{,l}$ with the PCAC
quark mass instead of the subtracted quark mass.

The reason for this behaviour has been traced back to the value of $\hopc$ taken as 
input from the literature. Indeed, for an accurate determination of $\hopc$ it
is crucial to follow a constant physics condition in the approach to the
continuum limit, which fixes the $\Oasq$ ambiguities coming from this source.
Instead, the value $\hopc=0.135496$ quoted in \cite{Rolf:2002gu} comes from an interpolation
of data obtained from a constant physics condition at other values of $\beta$.
While the effect of relaxing the constant physics requirement was found to
be negligible for the data of \cite{Rolf:2002gu}, its impact on the tuning
of twist angles is large. The $\beta=6.1$ critical point has been hence 
determined afresh, obtaining $\hopc=0.135665(11)$,
and $\beta=6.1$ simulations with new mass parameters have been performed.
Full details will be provided in a forthcoming publication \cite{cleanup}.

\begin{figure}
\vspace{65truemm}
\includegraphics{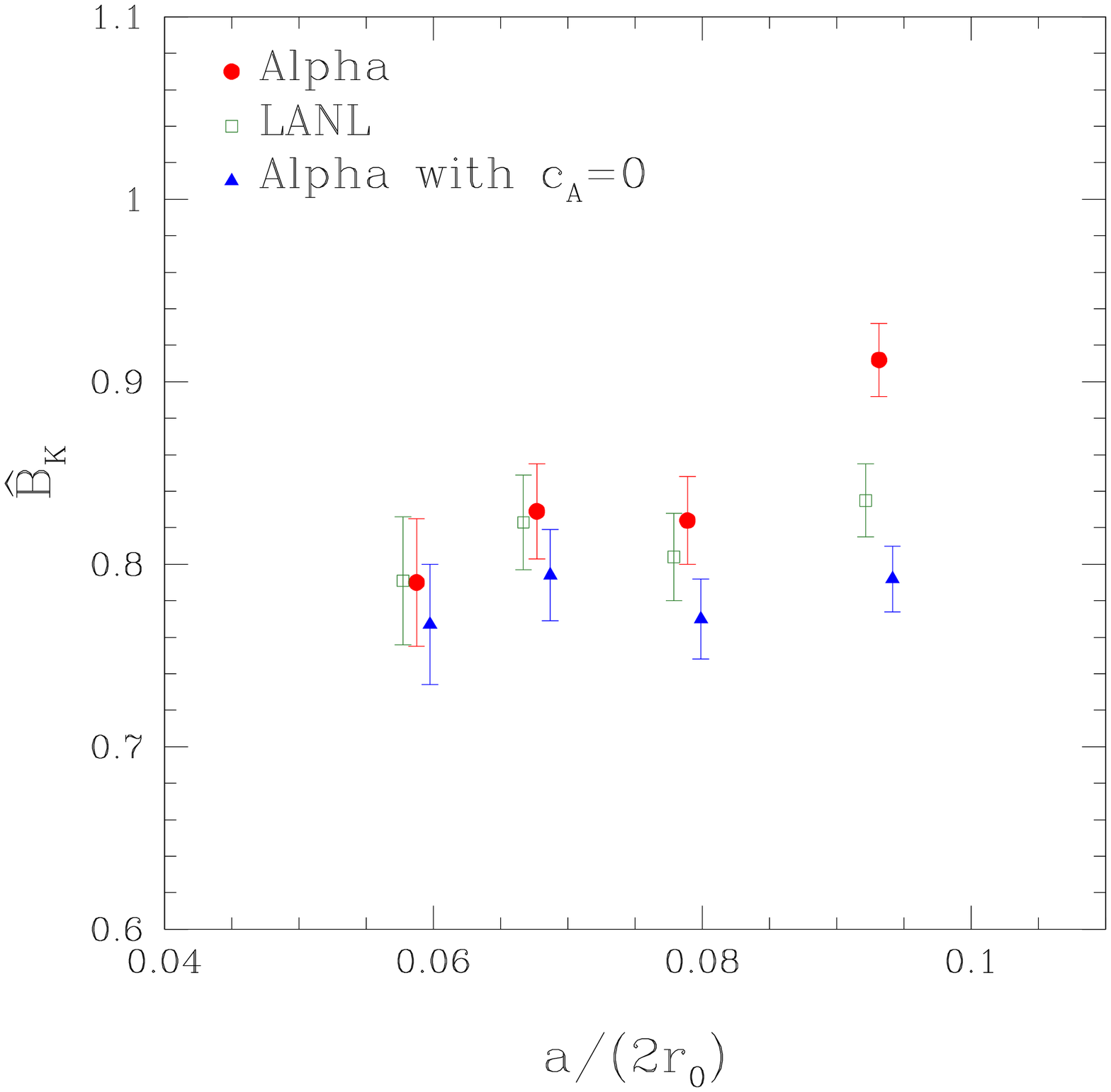}
\includegraphics{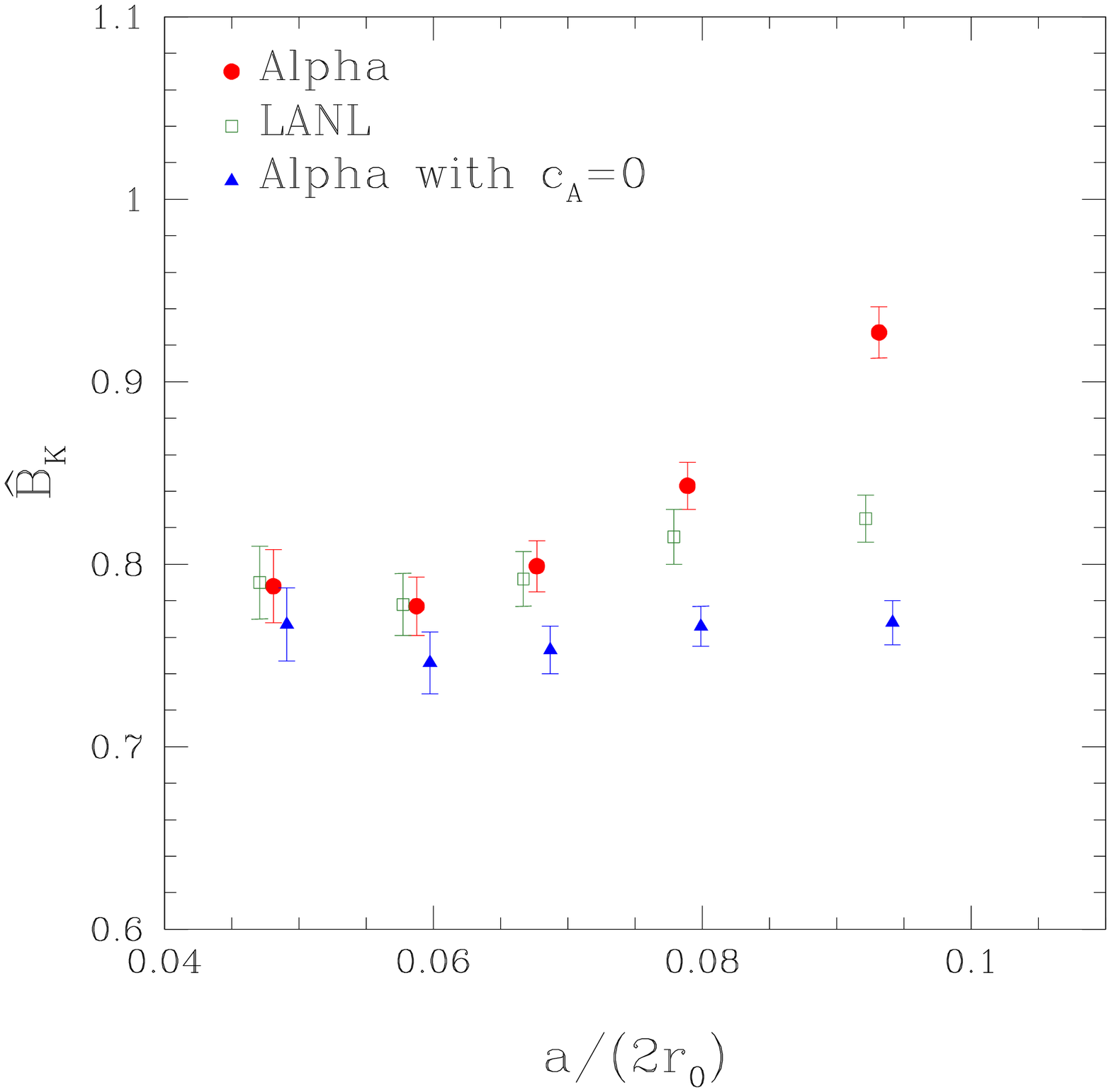}
\vspace{0truemm}
\caption{Uncertainties on $\hat B_K$ related to the $\Oa$ improvement of bilinears.
The left panel displays $\pi/2$ data, while the right panel shows $\pi/4$ data.}
\label{figimp}
\end{figure}

\subsection{Continuum limit}

As noted above, taking the continuum limit for $\hat B_K$ involves a linear 
extrapolation
in $a$. At this stage, having results from two different regularisations,
which can be combined in a fit constrained to a common continuum limit,
is essential for a proper control of the extrapolation.

\begin{figure}
\vspace{65truemm}
\includegraphics{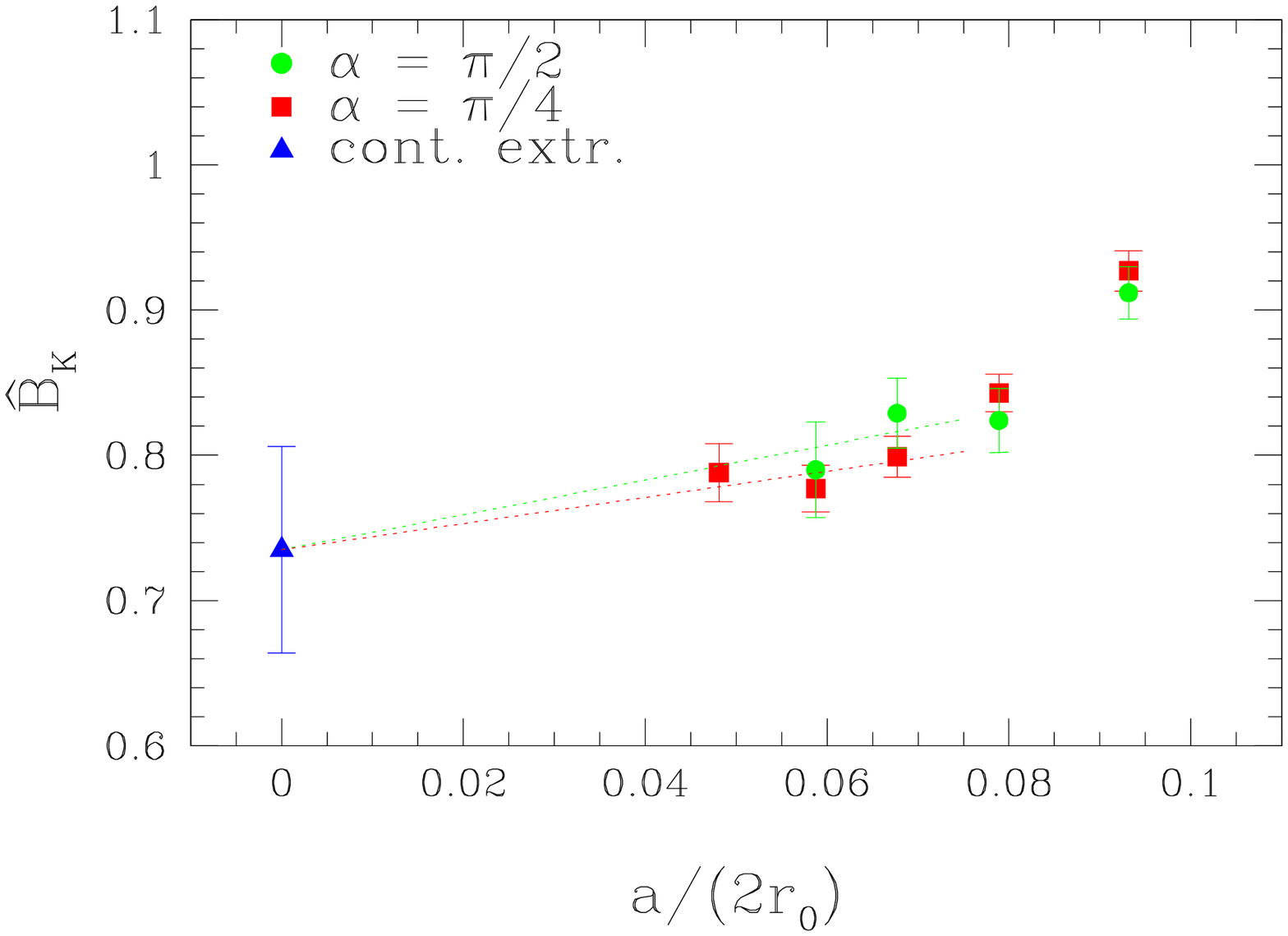}
\includegraphics{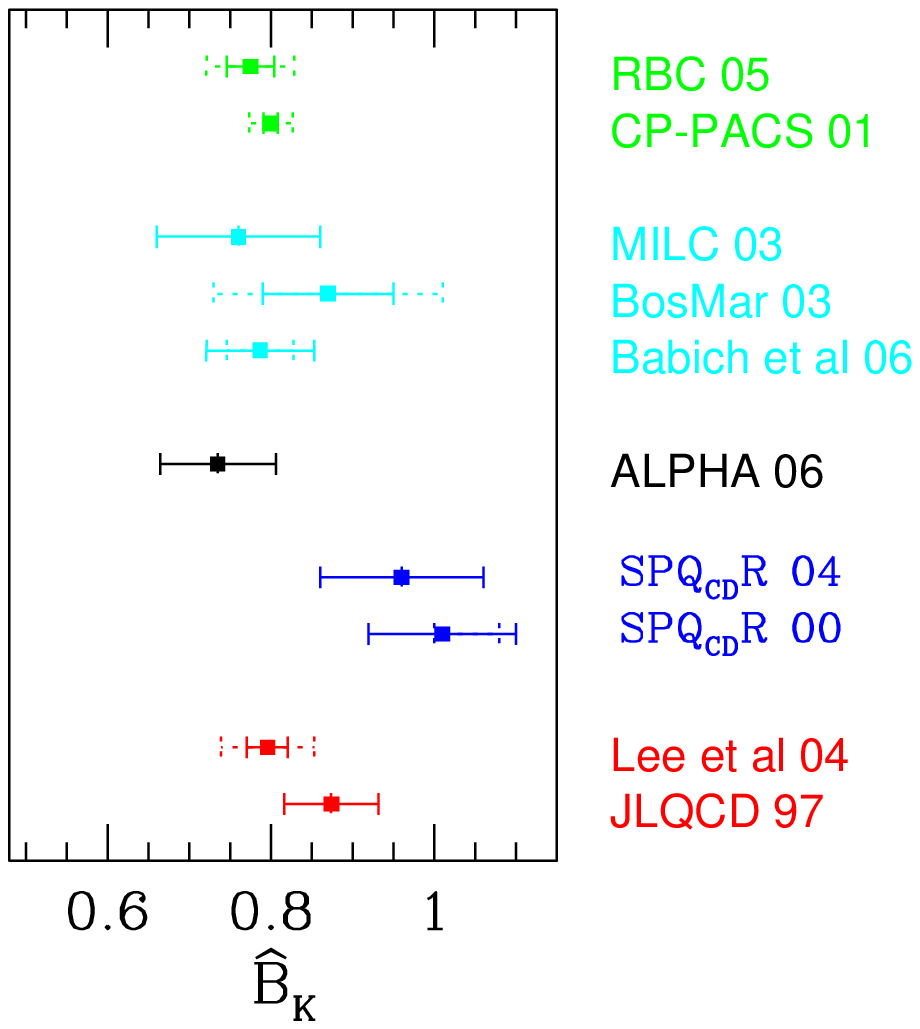}
\vspace{-5truemm}
\caption{Left: Continuum limit extrapolation of $\hat B_K$. Right: Comparison
with other quenched results.}
\label{fig2}
\end{figure}

It turned out that one of the most relevant sources of cutoff effects is
related to the construction of the $\Oa$ improved bilinears in the
denominator of Eq.~(\ref{eq:bare_ratio}). For instance,
using either the values for $\ZA,\ZV,\icA$ determined by the ALPHA Collaboration
or those obtained by the LANL group \cite{Bhattacharya:2001ks} results in sizeable effects
on $B_K$ at the lowest values of $\beta$ available (see Figure~\ref{figimp}). This
signals the presence of large $\Oasq$ ambiguities in $\hat B_K$ far from the
continuum limit. Combined linear+quadratic extrapolation of the data
proved to be unstable. Thus the values of $\beta$
for which the difference between ALPHA and
LANL constructions of $\Oa$ improved bilinears results to $\hat B_K$
discrepancies beyond one sigma were conservatively discarded in the linear
fits to the continuum limit. This means that results at
$\beta=6.0$ and $\beta=6.1$ had to be left out.
The resulting extrapolation is illustrated by the left panel of Figure~\ref{fig2}. 
The final results are:
\begin{gather}
\label{eq:BKRGI}
\hat B_K = 0.735(71) \,,\\
\bar B_K^{\MSbar}(2~\GeV) = 0.534(52) \,.
\end{gather}
When comparing with the result quoted in \cite{Dimopoulos:2006dm}, it has to
be taken into account that $\beta=6.1$
data have been revised, for the reasons explained above.

The value for $\hat B_K$ in Eq.~(\ref{eq:BKRGI}) is shown in
the right panel of Figure~\ref{fig2} alongside
other representative results in quenched QCD found in the literature. 
As discussed in
Appendix~E of \cite{Dimopoulos:2006dm}, the difference with other computations
with Wilson fermions is mainly due to the method employed to determine $B_K$:
instead of using a ratio similar to the one in Eq.~(\ref{eq:bare_ratio}),
the authors of \cite{SPQR:bk} extract $B_K$ from a fit of the mass
dependence of a different ratio of correlation functions, inspired by Chiral 
Perturbation Theory.
It has to be stressed that the computation of \cite{SPQR:bk} does not have direct
access to the physical kaon mass region.

The result of Eq.~(\ref{eq:BKRGI}) is the only existing quenched result
in the literature which
has simultaneously eliminated any systematic uncertainty related to
renormalisation (both at a reference scale and from the point of view
of RG running), ultraviolet cutoff dependences, and finite volume effects
(within the available accuracy). On the other hand, the control of the
mass dependence of $\hat B_K$ with Wilson fermions is still not as
accurate as with e.g. Neuberger or domain wall fermions. Overall, it seems
fair to claim that Eq.~(\ref{eq:BKRGI}) is a benchmark result for $\hat B_K$
in quenched QCD.

\section{A strategy to compute $B_B$}

Long-distance QCD contributions to indirect CP violation in the $B$-meson sector
of the SM are encoded in the bag parameters
\begin{gather}
\hat B_{B_\ell} = \frac{\langle \bar B^0|\hat O_\ell^{\Delta B=2}| B^0 \rangle}{\frac{8}{3}F_{B_\ell}^2 M_{B_\ell}^2} \, ,
\end{gather}
where the relevant effective four-quark interactions have the form
\begin{gather}
O_\ell^{\Delta B=2} = (\bar b \gamma_\mu^{\rm L} \ell)(\bar b \gamma_\mu^{\rm L} \ell) \, ,~~~~\ell=d,s \,.
\end{gather}
These B-parameters appear, together with the corresponding meson decay
constants, e.g. in the expressions for neutral $B$-meson mass differences
$\Delta M_\ell$ \cite{Battaglia:2003in}
\begin{align}
\Delta M_d &= 0.05~{\rm ps}^{-1}\,\times\,
\bigg[\frac{\sqrt{\hat B_{B_d}}F_{B_d}}{230~\MeV}\bigg]
\bigg[\frac{\bar m_t(\bar m_t)}{167~\GeV}\bigg]
\bigg[\frac{|V_{td}|}{0.0078}\bigg]
\bigg[\frac{\eta_B}{0.55}\bigg] \,,\\
\Delta M_s &= 17.2~{\rm ps}^{-1}\,\times\,
\bigg[\frac{\sqrt{\hat B_{B_s}}F_{B_s}}{260~\MeV}\bigg]
\bigg[\frac{\bar m_t(\bar m_t)}{167~\GeV}\bigg]
\bigg[\frac{|V_{ts}|}{0.040}\bigg]
\bigg[\frac{\eta_B}{0.55}\bigg] \,,
\end{align}
where $\eta_B$ encodes short-distance QCD effects.
The experimental values for these quantities are
$\Delta M_d=0.507\,\pm\,0.005~{\rm ps}^{-1}$ \cite{Yao:2006px} and
$\Delta M_s=17.77\,\pm\,0.10{\rm(stat)}\,\pm\,0.07{\rm(sys)}~{\rm ps}^{-1}$
\cite{Abulencia:2006ze}. The recent measurement of $\Delta M_s$ by the CDF 
Collaboration has set very stringent constraints
on the required precision of theoretical determinations of
$\hat B_{B_s}F_{B_s}^2$, which are now at the same level as
those on $\hat B_{B_d}F_{B_d}^2$. In addition to ``standard''
systematic uncertainties such as dynamical light quark effects,
matrix elements involving heavy quarks are particularly sensitive to
improvements coming from a systematic, conceptually controlled treatment
of heavy quark effects within lattice QCD.

A strategy for a precise lattice QCD computation of $B_{B_\ell}$,
which in principle would keep all systematic uncertainties under control,
has been put forward
in \cite{Palombi:2006pu}. The $b$ quark is treated at leading order in Heavy Quark
Effective Theory (i.e. in the static approximation), although $1/m_b$ corrections
from the heavy quark expansion can be eventually included following
the spirit of \cite{Heitger:2003nj}. In the static approximation the relevant
physical amplitude is a linear combination of matrix elements of two
static-light four-fermion operators, viz.
\begin{align}
\label{eq:opHL1}
O_1^{\Delta B=2} &= (\bar h \gamma_\mu^{\rm L} \ell)(\bar h \gamma_\mu^{\rm L} \ell) \, ,~~~~\ell=d,s \,,\\
\label{eq:opHL2}
O_2^{\Delta B=2} &= (\bar h (\Id-\gamma_5) \ell)(\bar h (\Id-\gamma_5) \ell) \, ,~~~~\ell=d,s \,.
\end{align}
The renormalisation of generic static-light four-fermion operators with
Wilson light fermions has been analysed in detail in \cite{Palombi:2006pu}. 
An important conclusion
of this study is that, similar to the case of fully relativistic operators, parity-even operators mix between them due to the
breaking of chiral symmetry. On the other hand, in the parity-odd sector it is possible to find a complete basis of
operators that renormalise multiplicatively. This opens the door to a generalisation
to this context of the tmQCD strategy pursued for $B_K$. In particular, it
is possible to extract $B_{B_\ell}$ from matrix elements
of the ${\rm VA+AV}$ and ${\rm SP+PS}$ parts of the operators in
Eqs.~(\ref{eq:opHL1},\ref{eq:opHL2}) if the light quark flavour $\ell$ is twisted
at $\alpha=\pi/2$. This avoids any need of dealing with complicated operator 
renormalisation patterns, and eliminates any constraint on quenched computations
due to exceptional configurations. Furthermore, it is possible to extend the
automatic $\Oa$ improvement arguments of Frezzotti and Rossi to show that the
matrix elements of interest will display scaling violations at $\Oasq$ only.

\begin{figure}
\vspace{80truemm}
\includegraphics{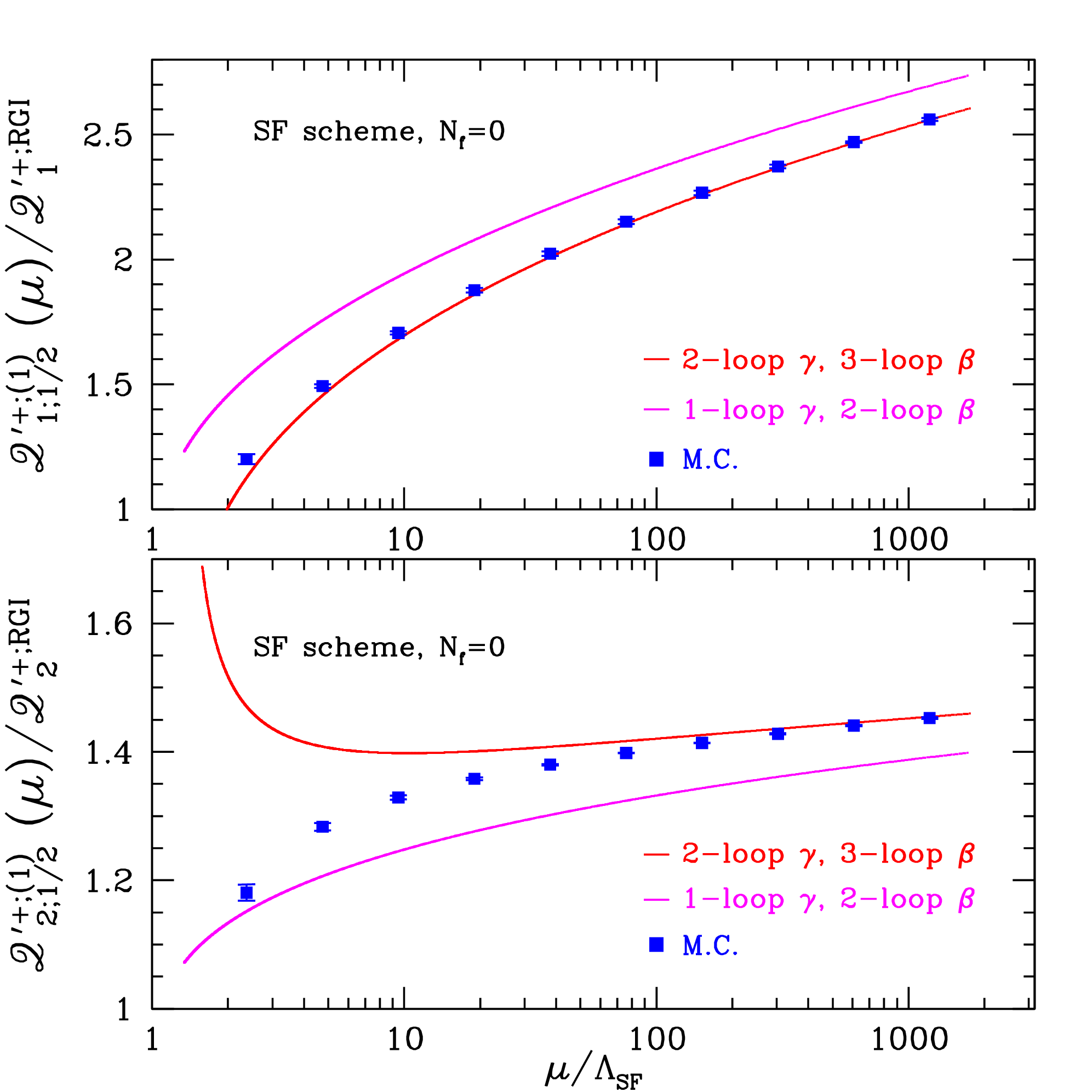}
\includegraphics{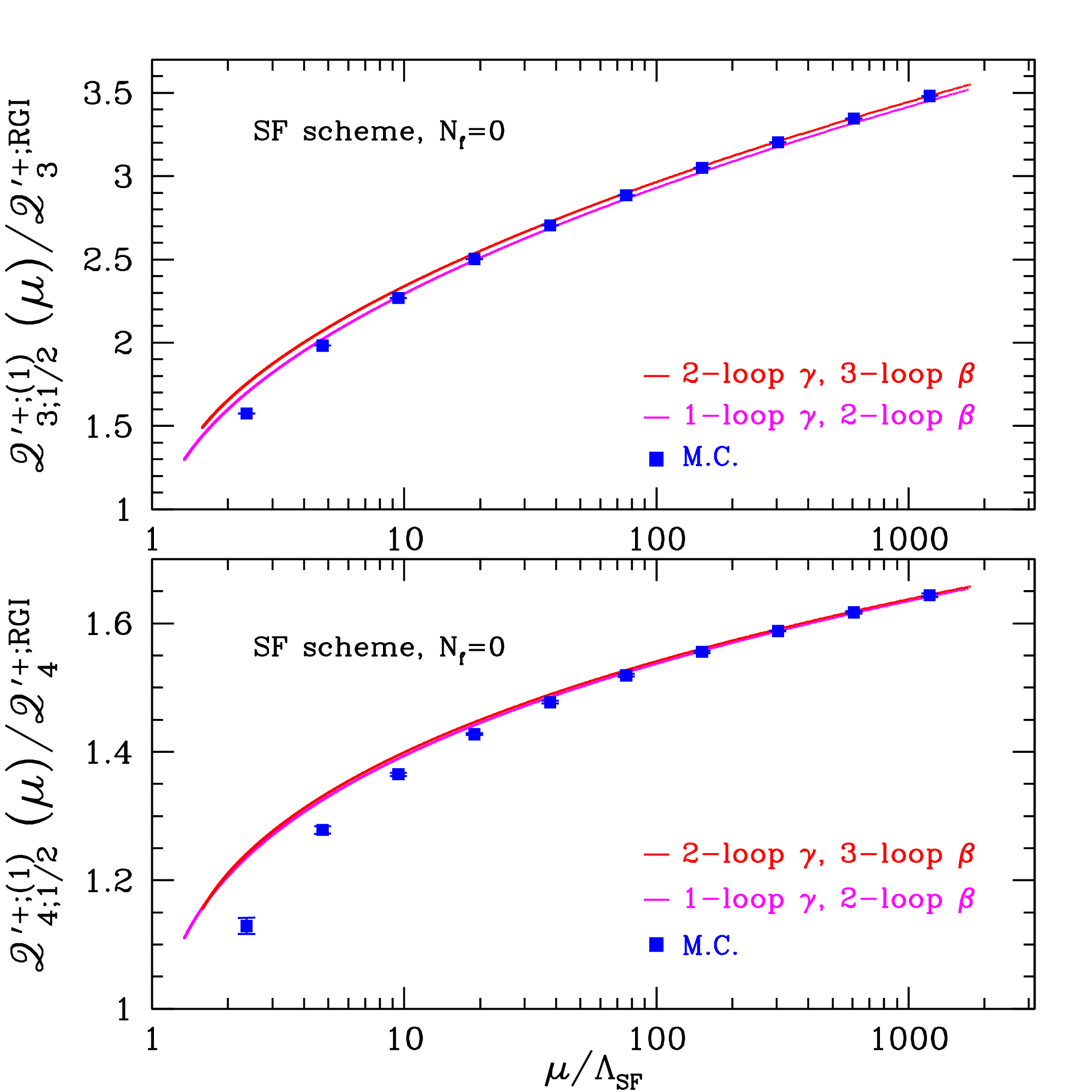}
\vspace{0truemm}
\caption{Preliminary results for the non-perturbative RG running (``M.C.'' points) of the 
four  independent elements of the multiplicatively renormalisable basis for parity-
odd static-light four-fermion operators. See \cite{Palombi:2006pu} for notational 
details. The solid curves are perturbative predictions.}
\label{fig3}
\end{figure}

The numerical implementation of non-perturbative renormalisation for
static-light four-fermion 
operators is discussed in detail in \cite{Palombi:2006pu}. Preliminary results for
the RG running of static-light four-fermion operators in quenched QCD are shown 
in Figure~\ref{fig3}.
Final results will be the object of a forthcoming publication \cite{BBNPR}.

\section{tmQCD for $K\to\pi\pi$?}

The computation of non-leptonic kaon decay amplitudes in QCD
poses much harder problems than those related to $\Delta F=2$ processes:

\begin{itemize}

\item Finite volume effects strongly affect the two-pion final state,
making the direct extraction of the amplitudes from Euclidean correlation
functions considerably difficult \cite{Maiani:1990ca,Lellouch:2000pv}. It has thus become customary
to attempt instead the computation of the relevant couplings in a low-
energy effective description of QCD based on Chiral Perturbation Theory, which in 
principle would allow the computation of the amplitudes at a given order in the 
chiral expansion \cite{Bernard:1985wf}. This requires, ideally, access to the chiral regime of QCD.

\item The renormalisation of the $\Delta S=1$ effective Hamiltonian arising
from the OPE treatment of electroweak interactions requires
dealing with a complex operator mixing problem. In particular, if chiral symmetry is 
not preserved by the lattice regularisation, as with Wilson fermions, mixing with 
lower  dimension operators proceeds via coefficients that diverge with an integer 
power of the cutoff \cite{Maiani:1986db}. On the other hand, if the charm quark
is kept as an active degree of freedom the presence of exact lattice chiral symmetry
eliminates all power-divergent mixings \cite{Capitani:2000bm,Giusti:2004an}.

\end{itemize}

The case for employing regularisations that preserve chiral symmetry in the approach
to this problem is therefore very strong. Indeed, chiral fermions have been
instrumental in a recent computation, in the quenched approximation,
of the leading-order low-energy couplings
of the $\Delta S=1$ effective weak Hamiltonian in the GIM limit $m_c=m_u$
\cite{Giusti:2006mh}, the first results of a
comprehensive programme aimed at understanding the r\^ole of the charm quark
in the $\Delta I=1/2$ enhancement rule \cite{Giusti:2004an,pilar}.
On the other hand, it is conceivable that the control over chiral symmetry
breaking that constitutes one of the main assets of tmQCD can be exploited
in order to alleviate the problems related to renormalisation. Two different
proposals have been actually put forward to that effect.

In \cite{Pena:2004gb}, a theory with four quark flavours is considered, without
specifying {\em a priori} how many of them are dynamical. Once the light
doublet is twisted at angle $\pi/2$, it is immediate to show that the power 
divergences
affecting $K\to\pi$ matrix elements are at most linear, and that there are
no finite mixings with other dimension six operators. This is a substantial
gain with respect to standard Wilson fermions, in which divergences are
quadratic and finite mixings are present. If the heavier $s,c$ flavours
are fully twisted as well (which is straightforward if they are kept quenched),
then it is possible to eliminate power divergences altogether, simply by
employing non-perturbatively $\Oa$ improved fermion action and quark bilinears.

In \cite{Frezzotti:2004wz}, the authors consider a theory in which a valence sector
containing an arbitrary number $N_v$ of flavours is matched to a theory
with $N_f$ dynamical quarks. All the flavours are fully twisted. The freedom
to fix twist angles arbitrarily for valence quarks, without the need
to restrict to non-anomalous chiral rotations, is then used to set up a
valence sector that allows to extract $K\to\pi$ matrix elements from correlation
functions that do not require any power divergent subtraction. The authors
propose a specific valence sector with $N_v=10$. In the same paper,
a similar technique is proposed to obtain a multiplicatively renormalisable
$B_K$; in this case, $N_v=6$. A strong advantage of this framework is that
only fully twisted quarks are used, and hence automatic $\Oa$ improvement
arguments apply.

These tmQCD proposals are appealing in that they potentially offer many of
the advantages of exactly symmetric regularisations at a considerably lower
computational cost. It has to be stressed, however, that the arguments which
show that undesired counterterms cancel rely crucially on the assumption
that a precise tuning of the twist angle has been performed. In the case of
power divergences the issue is particularly sensitive, as systematic uncertainties
in the tuning of parameters may result in a lack of cancellation of large
contributions to correlation functions. It is important to notice, too, that
the absence of exact chiral symmetry poses an intrinsic lower bound to the
quark masses that can be simulated safely; in particular, access to the deep
chiral regime, as achieved in \cite{Giusti:2006mh}, may be compromised. Finally,
the need to separate the $\Delta I=3/2$ and $\Delta I=1/2$ channels requires
a good control over the $\Oasq$ breaking of isospin symmetry inherent
to tmQCD.
Given these caveats, the suitability of tmQCD to deal with $K\to\pi\pi$ decays is an
open problem that may only be settled by dedicated numerical studies.

\section{Conclusions}

Twisted mass QCD, together with state-of-the-art techniques for Wilson fermions,
allow for benchmark quenched computations of weak matrix elements, as shown
by $B_K$. The ideas put forward for $\Delta S=2$ matrix elements can be
extended to other problems, like $\Delta B=2$ and $K\to\pi\pi$ amplitudes,
offering potential for precise computations that do not resort to exact
chiral symmetry.

The dominant source of uncertainty left in the quenched approximation
(certainly so for $B_K$) is related to the lack of full $\Oa$ improvement, which
amplifies the error of the continuum limit extrapolation. Thus, if Wilson
fermions are to be used in the future in the determination of weak matrix
elements, the use of tmQCD variants that embody automatic $\Oa$ improvement
\cite{Frezzotti:2004wz} may prove essential. Two important aspects of the tmQCD approach
are critical in the context of weak matrix elements: the tuning of
parameters, in particular of the twist angle, has to be controlled to high
precision; and flavour symmetry breaking effects have to be kept at the
few percent level. The question whether valence tmQCD quarks offer a convenient 
alternative to chirally symmetric fermions for some specific applications remains
to be addressed by dedicated simulations.

\section*{Acknowledgements}
I wish to thank my collaborators P.~Dimopoulos, M.~Guagnelli, J.~Heitger,
F.~Palombi, M.~Papinutto, S.~Sint, A.~Vladikas and H.~Wittig, as well as all
the other members of the ALPHA Collaboration, for several years of fruitful
work. I have enjoyed illuminating discussions on the topics covered in
this talk with many colleagues;
a special acknowledgement goes to D.~Be\'cirevi\'c, M.~Della~Morte, R.~Frezzotti, 
F.~Mescia, G.C.~Rossi and R.~Sommer. Finally, I wish to thank the staff at the Computing Center of
DESY-Zeuthen, whose unwavering technical support has been instrumental in most
of the numerical work described here.


\begin{thebibliography}{99}

\bibitem{Bona:2006ah}
  M.~Bona {\it et al.}  [UTfit Collaboration],
  arXiv:hep-ph/0606167.

\bibitem{LG}
L.~Giusti, PoS(LAT2006)009.

\bibitem{DelDebbio:2006cn}
  L.~Del Debbio, L.~Giusti, M.~L\"uscher, R.~Petronzio and N.~Tantalo,
  arXiv:hep-lat/0610059.

\bibitem{Jansen:2006rf}
  K.~Jansen and C.~Urbach  [ETM Collaboration],
  PoS(LAT2006)203
  [arXiv:hep-lat/0610015].

\bibitem{Gockeler:2006ns}
  M.~Gockeler {\it et al.}, 
  PoS(LAT2006)179
  [arXiv:hep-lat/0610066].

\bibitem{Gockeler:2006vi}
  M.~Gockeler {\it et al.},
  PoS(LAT2006)160
  [arXiv:hep-lat/0610071].

\bibitem{Kaneko:2006pa}
  T.~Kaneko {\it et al.}  [JLQCD Collaboration],
  PoS(LAT2006)054
  [arXiv:hep-lat/0610036].

\bibitem{Dimopoulos:2006dm}
  P.~Dimopoulos {\it et al.}
  ~[ALPHA Collaboration],
  Nucl.\ Phys.\ B 749 (2006) 69
  [arXiv:hep-ph/0601002].

\bibitem{Guagnelli:2005zc}
  M.~Guagnelli, J.~Heitger, C.~Pena, S.~Sint and A.~Vladikas  [ALPHA
                  Collaboration],
  JHEP 0603 (2006) 088
  [arXiv:hep-lat/0505002].

\bibitem{Palombi:2005zd}
  F.~Palombi, C.~Pena and S.~Sint,
  JHEP 0603 (2006) 089
  [arXiv:hep-lat/0505003].

\bibitem{Palombi:2006pu}
  F.~Palombi, M.~Papinutto, C.~Pena and H.~Wittig  [ALPHA Collaboration],
  JHEP 0608 (2006) 017
  [arXiv:hep-lat/0604014].

\bibitem{Lee:2006cm}
  W.~Lee,
  PoS(LAT2006)015
  [arXiv:hep-lat/0610058].

\bibitem{Onogi}
  T.~Onogi, these proceedings.

\bibitem{Yao:2006px}
  W.M.~Yao {\it et al.}  [Particle Data Group],
  J.\ Phys.\ G 33 (2006) 1.

\bibitem{Battaglia:2003in}
  M.~Battaglia {\it et al.},
  arXiv:hep-ph/0304132.

\bibitem{Marti:4fPT}
G.~Martinelli,
Phys. Lett. B141 (1984)
  395.

\bibitem{Bernardetal:4fPT}
C.~Bernard, T.~Draper, and A.~Soni,
Phys. Rev. D36 (1987) 3224.

\bibitem{Bernard:lat87}
C.~Bernard, T.~Draper, G.~Hockney, and A.~Soni,
  Nucl. Phys. Proc. Suppl. 4 (1998)
  483.

\bibitem{Guptaetal:bk}
R.~Gupta, D.~Daniel, G.~Kilcup, A.~Patel, and S.R.~Sharpe,
  Phys. Rev. D47 (1993) 5113
  [arXiv:hep-lat/9210018].

\bibitem{Doninetal:4fRIMOM}
A.~Donini, V.~Gim\'enez, G.~Martinelli, M.~Talevi, and A.~Vladikas,
Eur. Phys. J. C10 (1999) 121
  [arXiv:hep-lat/9902030].

\bibitem{BK:WI}
D.~{Be\'cirevi\'c} {\it et al.},
  Phys. Lett. B487 (2000) 74
  [arXiv:hep-lat/0005013].

\bibitem{SPQR:bk}
D.~{Be\'cirevi\'c}, P.~Boucaud, V.~{Gim\'enez}, V.~Lubicz, and M.~Papinutto,
  Eur. Phys. J. C37
  (2004) 315 [arXiv:hep-lat/0407004].

\bibitem{tmqcd:pap1}
 R.~Frezzotti, P.A. Grassi, S.~Sint, and P.~Weisz [ALPHA Collaboration],
  JHEP 08
  (2001) 058 [arXiv:hep-lat/0101001].

\bibitem{Babich:2006bh}
  R.~Babich {\it et al.}, 
  arXiv:hep-lat/0605016.

\bibitem{Luscher:1996ug}
  M.~L\"uscher, S.~Sint, R.~Sommer, P.~Weisz and U.~Wolff,
  Nucl.\ Phys.\ B 491 (1997) 323
  [arXiv:hep-lat/9609035].

\bibitem{Shindler:2005vj}
  A.~Shindler,
  PoS(LAT2005) 014
  [arXiv:hep-lat/0511002].

\bibitem{Frezzotti:2004pc}
  R.~Frezzotti,
  Nucl.\ Phys.\ Proc.\ Suppl.\  140 (2005) 134
  [arXiv:hep-lat/0409138].

\bibitem{Frezzotti:2003ni}
  R.~Frezzotti and G.C.~Rossi,
  JHEP 0408 (2004) 007
  [arXiv:hep-lat/0306014].

\bibitem{Sommer:2002en}
  R.~Sommer,
  Nucl.\ Phys.\ Proc.\ Suppl.\ 119 (2003) 185
  [arXiv:hep-lat/0209162].

\bibitem{Dimopoulos:2006es}
  P.~Dimopoulos {\it et al.}
  ~[ALPHA Collaboration],
  PoS(LAT2006)158
  [arXiv:hep-lat/0610077].

\bibitem{Dimopoulos:2006ma}
  P.~Dimopoulos {\it et al.},
  Phys.\ Lett.\ B 641 (2006) 118
  [arXiv:hep-lat/0607028].

\bibitem{Hernandez:2001yn}
  P.~Hern\'andez, K.~Jansen, L.~Lellouch and H.~Wittig,
  JHEP 0107 (2001) 018
  [arXiv:hep-lat/0106011].

\bibitem{Necco:2001xg}
  S.~Necco and R.~Sommer,
  Nucl.\ Phys.\ B 622 (2002) 328
  [arXiv:hep-lat/0108008].

\bibitem{Becirevic:2006ii}
  D.~Be\'cirevi\'c {\it et al.}, 
  Phys.\ Rev.\ D74 (2006) 034501
  [arXiv:hep-lat/0605006].

\bibitem{Rolf:2002gu}
  J.~Rolf and S.~Sint  [ALPHA Collaboration],
  JHEP 0212 (2002) 007
  [arXiv:hep-ph/0209255].

\bibitem{cleanup}
P.~Dimopoulos {\it et al.} [ALPHA Collaboration], in preparation.

\bibitem{Bhattacharya:2001ks}
  T.~Bhattacharya, R.~Gupta, W.~Lee and S.R.~Sharpe,
  Nucl.\ Phys.\ Proc.\ Suppl.\  106 (2002) 789
  [arXiv:hep-lat/0111001].

\bibitem{Abulencia:2006ze}
  A.~Abulencia {\it et al.}  [CDF Collaboration],
  arXiv:hep-ex/0609040.

\bibitem{Heitger:2003nj}
  J.~Heitger and R.~Sommer  [ALPHA Collaboration],
  JHEP 0402 (2004) 022
  [arXiv:hep-lat/0310035].

\bibitem{BBNPR}
  F.~Palombi, M.~Papinutto, C.~Pena and H.~Wittig  [ALPHA Collaboration], in preparation.

\bibitem{Maiani:1990ca}
  L.~Maiani and M.~Testa,
  Phys.\ Lett.\ B245 (1990) 585.

\bibitem{Lellouch:2000pv}
  L.~Lellouch and M.~L\"uscher,
  Commun.\ Math.\ Phys.\  219 (2001) 31
  [arXiv:hep-lat/0003023].

\bibitem{Bernard:1985wf}
  C.W.~Bernard, T.~Draper, A.~Soni, H.D.~Politzer and M.B.~Wise,
  Phys.\ Rev.\ D32 (1985) 2343.

\bibitem{Maiani:1986db}
  L.~Maiani, G.~Martinelli, G.C.~Rossi and M.~Testa,
  Nucl.\ Phys.\ B289 (1987) 505.

\bibitem{Capitani:2000bm}
  S.~Capitani and L.~Giusti,
  Phys.\ Rev.\ D64 (2001) 014506
  [arXiv:hep-lat/0011070].

\bibitem{Giusti:2004an}
  L.~Giusti, P.~Hern\'andez, M.~Laine, P.~Weisz and H.~Wittig,
  JHEP 0411 (2004) 016
  [arXiv:hep-lat/0407007].

\bibitem{Giusti:2006mh}
  L.~Giusti {\it et al.}, 
  arXiv:hep-ph/0607220.

\bibitem{pilar}
  P.~Hern\'andez, these proceedings.

\bibitem{Pena:2004gb}
  C.~Pena, S.~Sint and A.~Vladikas,
  JHEP 0409 (2004) 069
  [arXiv:hep-lat/0405028].

\bibitem{Frezzotti:2004wz}
  R.~Frezzotti and G.C.~Rossi,
  JHEP 0410 (2004) 070
  [arXiv:hep-lat/0407002].

\end{thebibliography}
\end{document}